\documentclass[aps,prb,
                twocolumn,
                floatfix,
                showpacs,
                amsmath,
                amssymb]{revtex4}

\usepackage{graphicx}
\usepackage{bm}

\DeclareMathOperator{\Trres}{Tr_{\mit{res}}}
\DeclareMathOperator{\Trdot}{Tr_{\mit{dot}}}
\newcommand{\nts}{\negthickspace}

\newcommand{\sizeiii}{\scriptsize}

\newcommand{\ve}[1]{\mathbf{#1}}

\newcommand{\ket}[1]{\left\vert #1 \right \rangle}
\newcommand{\bran}[1]{\langle #1 \lvert}
\newcommand{\ketn}[1]{\rvert #1 \rangle}
\newcommand{\braket}[2]{\langle #1\mspace{1.2mu}\vert\mspace{1.2mu}#2\rangle}
\newcommand{\abs}[1]{\lvert #1\rvert}
\newcommand{\absbig}[1]{\bigl\vert #1\bigr\vert}
\newcommand{\imag}{\textup{i}}
\newcommand{\commutator}[2]{\mathopen[#1,#2\mathclose]_{\scriptscriptstyle -}}
\newcommand{\iop}{\textup{i} 0^{\!\scriptscriptstyle+}\!}
\newcommand{\Jdot}{J_{\mspace{-1mu}\mit{dot}}}
\newcommand{\Bzee}{B_{\mspace{-1mu}\text{\textsl{Zee}}}}
\newcommand{\Ham}{\mathcal{H}}
\newcommand{\Hdot}{\mathcal{H}_{\mspace{-1mu}\mit{dot}}}
\newcommand{\pst}{p^{\text{\slshape st}}}
\newcommand{\dH}{\Delta\mspace{-2.0mu}H}
\newcommand{\LiouvO}{\textup{L}_0}
\newcommand{\LiouvT}{\textup{L}_T}
\newcommand{\rhoeqres}{\rho_{\mit{res}}^{\text{\slshape eq}}}
\newcommand{\To}{\mathsf{T_{\!0}}}
\newcommand{\So}{\mathsf S}
\newcommand{\Td}{\mathsf{T_{\!-}}}
\newcommand{\Tu}{\mathsf{T_{\!+}}}
\newcommand{\udo}{\uparrow\downarrow}
\newcommand{\duo}{\downarrow\uparrow}
\newcommand{\s}{{s^{\vspace{0.5ex}}}}
\newcommand{\si}{{s^{\vspace{0.5ex}}_1}}
\newcommand{\sip}{{s'_1}}
\newcommand{\sii}{{s^{\vspace{0.5ex}}_2}}
\newcommand{\siip}{{s'_2}}
\newcommand{\SErq}{\Sigma^{r\mspace{1mu}q}}
\newcommand{\deltassp}{\Delta_{\mspace{-1mu}s^{\vspace{0.5ex}}s'}}
\newcommand{\deltasps}{\Delta_{\mspace{-1mu}s's^{\vspace{0.5ex}}}}
\newcommand{\deltaSTd}{\Delta_{\mathsf{S}\mathsf{T_{\!-}}}}
\newcommand{\sijrp}[4]{\sigma_{#1#2}^{#3\,#4}}

\begin{document}


\title{Magnon transport and spin current switching through quantum dots}

\author{Olaf Strelcyk}
\author{Thomas Korb}
\author{Herbert Schoeller}
\affiliation{Institut f\"ur Theoretische Physik, Rheinisch-Westf\"alische Technische Hochschule Aachen, 52056 Aachen, Germany}

\date{\today}

\begin{abstract}
We study the nonequilibrium spin current through a quantum dot consisting of two localized spin\nobreakdash-$1/2$ coupled to two ferromagnetic insulators. The influence of an intra-dot magnetic field and exchange coupling, different dot-reservoir coupling configurations, and the influence of magnon chemical potential differences vs.~magnetic field gradients onto the spin current are examined. We discuss various spin switching mechanisms and find that, in contrast to electronic transport, the spin current is very sensitive to the specific coupling configuration and band edges. In particular, we identify 1- and 2-magnon transport processes which can lead to resonances and antiresonances for the spin current.
\end{abstract}

\pacs{75.10.Jm, 75.40.Gb, 75.75.+a, 73.63.Kv}


\maketitle

%
\section{Introduction}
Electronic transport in mesoscopics has been widely studied for the last two decades. Particularly due to the interplay between the electronic charge and spin many novel features have been found, which lead to new inventions as giant magnetoresistance, tunneling magnetoresistance or the  Datta-Das transistor (see e.g. Rfs.~\onlinecite{prinz, awschalom, wolf,maekawa_shinjo} for reviews). This has opened the field of spintronics with possible future applications in the field of quantum computing. The fact that spin relaxation and dephasing rates are rather long compared to charge excitations life-times has recently stimulated the investigation of spin transport in purely magnetic systems without an accompanying charge transport. Experimentally, anomalous heat transport has been reported in $(Sr, Ca, La)_{14}Cu_{24}O_{41}$-materials \cite{experiment}, which is explained by the specific dynamics of spin excitations in quasi-onedimensional systems \cite{theory}. These experiments are still in the incoherent regime where the mean free path of the spinons is much smaller than the system size. In contrast, Meier and Loss\cite{meierloss} investigated coherent spin transport through mesoscopic ferromagnetic and antiferromagnetic Heisenberg spin chains. For the case of the isotropic antiferromagnetic spin\nobreakdash-$1/2$ chain they found that the spin conductance is quantized in units of order $(g\,\mu_{\rm  B})^2/h$. Matveev\cite{matveev} studied spinon transport through a one-dimensional Wigner crystal, which effectively can be described by an antiferromagnetically ordered spin chain. Here, anomalous conductance quantization phenomena occur which might explain experiments through quantum point contacts \cite{thomas}.\\
Motivated by these interesting recent results for spin transport through one-dimensional spin chains, we analyze in this work the magnetic analog of electronic transport through zero-dimensional quantum dots in the weak coupling regime (i.e.~weak exchange coupling between the magnetic quantum dot and the magnetization reservoirs). In particular, we study the spin current through a quantum dot consisting of two interacting spin\nobreakdash-$1/2$, coupled to two ferromagnetic insulators, which are for simplicity described by a ferromagnetic Heisenberg model (the basic transport features revealed here are expected to hold for any reservoir involving magnon-like excitations). For the same system Wang et al.\cite{wang} derived a Landauer-B\"uttiker-type formula for spin current transport. However, they studied the magnon transport using a magnon representation for the dot states and treated the intra-dot magnon-magnon interaction in a mean-field approximation. In contrast, we study the case of arbitrarily strong intra-dot interaction and, using a real-time formalism developed in Ref.~\onlinecite{schoellerHab}, perform only a perturbative expansion in the exchange coupling between the quantum dot and the magnetization reservoirs. This is the regime where in electronic transport interesting effects like Coulomb blockade phenomena occur.\\ 
Our main focus lies on the investigation of possibilities to switch the spin current by tuning system parameters as e.g.~an intra-dot magnetic field or the exchange coupling between the two spin\nobreakdash-$1/2$. In Rfs.~\onlinecite{burkardloss,burkardseelig} it was shown that the latter may be modulated, e.g. by application of gates. In contrast to electronic transport, we find that the specific microscopic coupling configuration and the magnon dispersion relation in the reservoirs has a strong influence on the spin current. Interestingly, the spin current can not only be dominated by 1-magnon transport processes, but also by 2-magnon cycles, leading to specific resonances or antiresonances in the spin current as function of the intra-dot exchange coupling. 
Furthermore, we discuss two different mechanisms of driving the spin current, firstly a difference of left and right magnon chemical potentials, and secondly a magnetic field gradient. We find that they are not equivalent but yield different results, unlike the counterpart of electrochemical potentials in electronic transport.\\
The paper is organized as follows. In section~\ref{sec:model} we introduce the model and the extension of the real-time formalism to bosonic transport. Assuming the reservoirs to be magnetized in the same direction, subsequently, in section~\ref{sec:deltamu} the spin current is discussed for the case of differing chemical potentials, in section~\ref{sec:deltaB} for the case of a magnetic field gradient applied, in each case for two different dot-reservoir coupling configurations.
In section~\ref{sec:antiparallel} we deal with the case of antiparallel magnetized reservoirs and finally close with some continuative remarks concerning extensions to this work. 
\section{Model\label{sec:model}}
Our model consists of a quantum dot, coupled via local exchange to two magnetization reservoirs. The dot is made up of two coupled spin\nobreakdash-$1/2$, whereas the reservoirs are assumed to be ferromagnetic insulators, which can be described by a ferromagnetic Heisenberg model, see Fig.~\ref{fig:coupleconfigs}.\\
The total Hamiltonian $\bar\Ham$ for the system can be written as sum of the dot Hamiltonian, the reservoir contributions and the exchange part, describing the coupling between the dot and the reservoirs (in the following called tunneling part corresponding to tunneling of magnons instead of electrons in the electronic analog):
\begin{equation}
\bar\Ham = \bar\Ham_{\mspace{-1mu}\mit{dot}} + \bar\Ham_L + \bar\Ham_R + \bar\Ham_T\,.
\end{equation}
The dot Hamiltonian $\bar\Ham_{\mspace{-1mu}\mit{dot}}$ reads 
\begin{equation}
\bar\Ham_{\mspace{-1mu}\mit{dot}} = -\Jdot\,\ve{s}_1\ve{s}_2 + \Bzee\,s^z,
\end{equation}
where $\Jdot$ is the exchange coupling between the two spin\nobreakdash-$1/2$, $s^z = s_1^z +s_2^z$ and $\Bzee$ represents a homogeneous magnetic field applied on the dot.\footnote
{
The factor $g\mu_{\rm B}$ is included in the Zeeman energy $\Bzee$, in the following referred to as a magnetic field. This is also the case with all following magnetic fields, resp.~Zeeman energies.
}
The reservoir contributions are given by a Heisenberg Hamiltonian:
\begin{equation}
\bar\Ham_r = -J/2 \sum_{\langle i,j\rangle}\ve{S}_{ri}\ve{S}_{rj} + B_r\sum_i S_{ri}^z + b_1 S_{r1}^z\,,
\end{equation}
with $r\in\{L,R\}$, $J>0$ and $\langle i,j\rangle$ denoting the sum over neighboring sites.\footnote
{
In this work we assume one-dimensional reservoirs. However, since the qualitative shape of the spectral function does not depend on the reservoir dimension, apart from minor modulations, no qualitative changes occur for reservoirs of other dimensionality.
}
$B_r$ represents a homogeneous magnetic field applied on the reservoir~$r$, whereas $b_1$ models a remnant of the dot magnetic field $\Bzee$. For simplicity we assume that this remnant field acts exclusively on the reservoir sites neighboring the dot.\footnote
{
The consideration of the remnant field $b_1$ is necessary to break isotropy. This is due to a subtlety occurring in the calculation of the spectral function $\Gamma(\omega)$ [cf.~\eqref{eq:specfunc}] for the semi-infinite Heisenberg model. As a consequence of this $b_1 = 0$ represents an unphysical border case.\\
Isotropy may also be broken by considering an anisotropic Heisenberg model (z-anisotropy). However, as long as a ferromagnetic ground state prevails, this does not lead to qualitative modifications of the results presented here.
}
The tunneling part is given by
\begin{equation}\label{eq:barHamT}
\bar\Ham_T = -\nts\sum_{r=L,R}\,\sum_{i=1,2}J^i_r\: \ve{s}_i\,\ve{S}_{r1}\,\,.
\end{equation}
where $J^i_r$ are the real exchange couplings between the two dot spins and the adjacent spins of the reservoirs, in the following referred to as tunnel couplings.\\
For the case of parallel magnetized reservoirs we assume the reservoir ground states to be $\ket{S_i^z = -S\,,\; \forall\,i}$ (the modifications for the case of antiparallel magnetized reservoirs will be stated in section~\ref{sec:antiparallel}).  Making use of the Holstein-Primakoff transformation, $\bar\Ham_r$ may be diagonalized by introducing the magnon creation (annihilation) operators $a^\dagger_{rk}$ ($a_{rk}$):
\begin{equation}\label{eq:Ham_r}
\Ham_r = \sum_k \omega_{rk}\, a^\dagger_{rk}a_{rk}, \quad \omega_{rk} = 2JS[1-\cos(ka)] + B_r, 
\end{equation}
with the lattice constant~$a$. Here and throughout the paper we set $\hbar = k_{\rm B} = 1$. Furthermore, in the following all energies as also the spin current are normalized to $JS$.\\ 
Rewriting $\bar\Ham_T$ in terms of the reservoir magnon creation and annihilation operators yields
\begin{equation}\label{eq:HamT}
\Ham_T = \sum_{r,i,k} \left( J^i_{rk}\: s_i^+ a_{rk} + \text{\,\it h.\,c.}\right),
\end{equation}
with the $k$-dependent coupling $J^i_{rk} \propto -J^i_r\, \braket{1}{k}$, where $\braket{1}{k}$ represents the amplitude of the magnon wavefunction at the dot neighboring reservoir sites. 
In~\eqref{eq:HamT} we have split off the ${s}_i^z\,{S}_{r1}^z$-term arising in~\eqref{eq:barHamT} and have included its ground state average (i.e. replacing $S^z_{r1}\rightarrow -S$) into the dot Hamiltonian.
This leads to an effective field $H_i = \sum_r J^i_r\:S$ for the dot states:
\begin{equation}
\Hdot = -\Jdot\,\ve{s}_1\ve{s}_2 + \Bzee\, s^z + \left( H_1\,s_1^z + H_2\,s_2^z \right).
\end{equation}
Thus our final Hamiltonian $\Ham$ reads
\begin{equation}
\Ham = \Ham_0 + \Ham_T\,,\quad\Ham_0 = \Ham_L + \Ham_R + \Hdot\,.
\end{equation}
The triplet states $\ket{\Tu} = \ket{\uparrow\uparrow}$ and $\ket{\Td} = \ket{\downarrow\downarrow}$ are eigenstates of $\Hdot$.
Within the sub-space spanned by $\ket{\So} = \tfrac{1}{\sqrt{2}}\left(\,\ket{\udo} - \ket{\duo}\,\right)$ and $\ket{\To} = \tfrac{1}{\sqrt{2}}\left(\,\ket{\udo} + \ket{\duo}\,\right)$, i.e.~the states with $s^z$ quantum number $m=0$, $\Hdot$ corresponds to the matrix
$\bigl[\begin{smallmatrix}\Jdot& \dH\\ \dH& 0\end{smallmatrix}\bigr]$. $\dH = (H_1 - H_2)/2$ measures the inhomogeneity of the effective field $H_i$, which originates from the tunnel couplings.

We consider two different mechanisms driving the spin current through the dot, which we denote as $\Delta\mu$- and \emph{$\Delta B$-configuration} in the following.\\
For the \emph{$\Delta\mu$-configuration} the magnetization of the reservoirs, given by the magnon number, is supposed to be conserved. This means that we assume the spin relaxation time in the reservoirs to be longer than the typical spin current measurement time. Therefore a chemical potential $\mu_r$ is assigned to the magnons in each of the reservoirs. Moreover we set $B_r = 0$ here, so that only the difference between the magnon chemical potentials $\mu_L$ and $\mu_R$ drives the spin current.\\
For the \emph{$\Delta B$-configuration} we set $\mu_r = 0$ in both reservoirs. Here a magnetic field gradient $B_{L/R}(t) = B_0 \pm \Delta B\,$, switched on simultaneously with the tunneling, gives rise to a non-vanishing spin current. The offset field $B_0 \geq \abs{\Delta B}$ introduced here is required for the stabilization of the ferromagnetic ground state.
%

We use the real-time formalism described in Ref.~\onlinecite{schoellerHab} to compute the stationary spin current. By integrating out the reservoir degrees of freedom this formalism yields an effective description of the system in terms of the dot degrees of freedom. It is based on the formally exact kinetic equations (tunneling and magnetic field gradient are assumed to be switched on at time $t=0$)
\begin{gather}
I_r(t) = \Trdot\Bigl[\, \int_{0}^{t}\nts dt'\;\Sigma_{I_r}(t-t')\,p(t') \Bigr],\\
\dot{p}(t) = -\imag\, \LiouvO\,p(t) + \int_{0}^t \nts dt'\; \Sigma(t-t')\,p(t'),
\end{gather}
where $I_r(t)$ denotes the time-dependent expectation value of the spin current through the dot. The current operator $\hat{I}_r$ is given by
\begin{equation}\label{eq:hatI}
\hat{I}_r = -\imag\,\commutator{\Ham}{N_r} =\imag \sum_{i,k} \left( J^{i\:*}_{rk}\: a^\dagger_{rk}s_i^- - J^i_{rk}\: s_i^+ a_{rk} \right)\,
\end{equation}
with $\commutator{\,\cdot}{\cdot\,}$ denoting the commutator and $N_r = \sum_k a^\dagger_{rk}a_{rk}$, so that the spin current $I_r$ is positive, when a `spin\nobreakdash-up' leaves the reservoir~$r$.
$p(t)$ is the time-dependent reduced density matrix, which is the trace over the reservoir degrees of freedom, $\Trres$, of the full density matrix. In contrast, $\Trdot$ denotes the trace over the local (dot) degrees of freedom. The free propagation of the system is determined by the Liouvillian $\LiouvO$. It is defined by $\LiouvO = \commutator{\Ham_0}{\cdot\,}\,$. 
The coupling to the reservoirs is described by the integral kernels $\Sigma$ and $\Sigma_{I_r}$. They are given by
\begin{gather}
\begin{split}
\Sigma&(t-t') \\ 
&= (-\imag)^2\!\Trres\!\!\left[ \LiouvT e^{-\imag\LiouvO t} T e^{-\imag \int_{t'}^t \!d\tau \LiouvT(\tau)} e^{\imag \LiouvO t'} \LiouvT \rhoeqres  \right]_{\text{\textsl{irred.}}}\!,
\end{split}\\
\begin{split}
\Sigma_{I_r}(t-t')
=& -\imag\Trres\!\Bigl[ \textup{A}_{I_r}e^{-\imag\LiouvO t} T e^{-\imag \int_{t'}^t \!d\tau \LiouvT(\tau)}\\ 
& \times e^{\imag \LiouvO t'} \LiouvT \rhoeqres  \Bigr]_{\text{\textsl{irred.}}},
\end{split}
\end{gather}
where we assumed that the reservoirs are in equilibrium described by the reservoir density matrix $\rhoeqres$. $\LiouvT = \commutator{\Ham_T}{\cdot\,}$ is the interaction part of the Liouvillian, and the interaction picture is defined by $\LiouvT(t) = e^{\imag\LiouvO t} \LiouvT e^{-\imag\LiouvO t}$. The superoperator $\textup{A}_{I_r}$ is given by the anticommutator with the current operator $\hat{I}_r$:
\begin{equation}
\textup{A}_{I_r} = \tfrac{1}{2}\mathopen[\hat{I}_r,\cdot\,\mathclose]_{\scriptscriptstyle +}\,.
\end{equation}
$T$ denotes the time ordering operator, and the index ``\textsl{irred.}'' indicates that only irreducible diagrams are taken into account (meaning that each vertical cut through a diagram will cut at least one bosonic reservoir line, see Ref.~\onlinecite{schoellerHab} for details).
Introducing the Laplace transforms $f(z) = \int_0^\infty dt e^{\imag z t} f(t)$ of the time-dependent functions $f(t)$ and using the identity $\lim_{t\rightarrow \infty}{f(t)} = -\imag\lim_{z\rightarrow\iop} zf(z)$ leads to the following expression for the stationary spin current $I_r^{\text{\slshape st}}$:
\begin{equation}\label{eq:Irst}
I_r^{\text{\slshape st}} = \Trdot{\Bigl[ \Sigma_{I_r}(\iop)\,\pst  \Bigr]}\,,
\end{equation}
where the stationary reduced density matrix $\pst$ is determined by 
\begin{equation}\label{eq:pst}
\Bigl[ -\imag\,\LiouvO + \Sigma(\textup{i} 0^{\!\scriptscriptstyle+}\!) \Bigr]\, \pst = 0.
\end{equation}
We calculate the kernels $\Sigma$ and $\Sigma_{I_r}$ to lowest order in tunneling, i.e.~second order in the coupling constants $J^i_r$. The diagrams contributing are listed in the appendix. 
A typical diagram is shown in Fig.~\ref{fig:diagram}(a), which corresponds to the following expression:
\begin{figure}
\includegraphics[width=\linewidth]{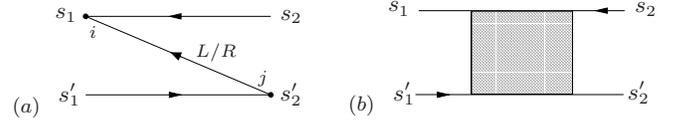}
\caption{\label{fig:diagram}
$(a)$ A typical diagram contributing to the kernels $\Sigma$ and $\Sigma_{I_r}$.
$(b)$ The hatched block represents arbitrary contractions.}
\end{figure}
\begin{equation}\label{eq:diagexpression}
\begin{split}
\frac{\imag}{2\pi} \sum_{i,j}\bran{\si} J^i_{\scriptscriptstyle L/R} & s_i^+ \ketn{\sii} \bran{\siip} J^j_{\scriptscriptstyle L/R} s_j^- \ketn{\sip}\\
& \times \int d\omega \frac{\Gamma(\omega)\, n_{\scriptscriptstyle L/R}(\omega)}{\Delta_{\sip\sii}\mp \Delta B - \omega +\iop}\,,
\end{split}
\end{equation}
where $\Delta_{\sip\sii} =E_{\sip}-E_{\sii}$ and $E_s$ are the eigenvalues of $\Hdot$. $n_r(\omega) = 1/\bigl(e^{\beta (\omega - \mu_r)} -1\bigr)$ is the Bose function and the spectral function $\Gamma(\omega)$ is given by
\begin{equation}\label{eq:specfunc}
\Gamma(\omega) =  \pi S \sum_k \abs{\braket{1}{k}}^2\, \delta(\omega-\omega_{k}),
\end{equation}
with $\omega_k = 2JS[1-\cos(ka)] + B_0$. In~\eqref{eq:diagexpression} $\mp\Delta B$ corresponds to the left/right reservoir. 
Due to the conservation of the total spin the following equation holds for the $s^z$ quantum numbers of all non-vanishing diagrams, as indicated in Fig.~\ref{fig:diagram}(b):
\begin{equation}\label{eq:mmmm}
m_{\si} - m_{\sip} = m_{\sii} - m_{\siip}\,.
\end{equation}
Assuming the dot to be in thermal equilibrium initially, the non-diagonal matrix elements $p(0)_{s,s'}$ of the initial reduced density matrix vanish.
From~\eqref{eq:mmmm} it then follows that the only non-trivial non-diagonal elements of the reduced density matrix $p(t)$ (particularly $\pst$) generated are those between the two states with $m = 0$.\\
In some cases the influence of these non-diagonal elements onto the stationary spin current may be neglected. Then~\eqref{eq:Irst} and~\eqref{eq:pst} reduce to classical master equations:
\begin{gather}
\sum_{r,s'}\sum_{q=\pm} \Bigl( \SErq_{\s,s'}\: \pst_{s'} - \SErq_{s',\s}\: \pst_{\s}\Bigr) = 0\,,\\
I_r^{\text{\slshape st}} = \sum_{\s,\,s'} \left( \Sigma^{r+}_{\s, s'} \:\pst_{s'} - \Sigma^{r-}_{s',\s}\:\pst_\s \right),\label{eq:classIrst}
\end{gather}
where $q=\pm$ corresponds to transitions with a spin-flip of $\Delta m = \pm 1$ in the dot.
The rates $\Sigma^{r\pm}_{\s,s'}$ are given by
\begin{subequations}\label{eq:SELR}
\begin{align}
\Sigma^{{\scriptscriptstyle L/R}+}_{\s, s'}& = \absbig{\bran{s} \textstyle\,\sum_i J_{\scriptscriptstyle L/R}^i s_i^+ \ketn{s'}}^2\notag\\
        &\quad\quad\times\Gamma(\deltassp \mp\Delta B)\,n_{\scriptscriptstyle L/R}(\deltassp \mp\Delta B),\label{eq:SELRp}\\ 
\Sigma^{{\scriptscriptstyle L/R}-}_{\s, s'}& = \absbig{\bran{s'} \textstyle\,\sum_i J_{\scriptscriptstyle L/R}^i s_i^+ \ketn{s}}^2\notag\\
        &\quad\quad\times\Gamma(\deltasps \mp\Delta B)\,\bigl[1+n_{\scriptscriptstyle L/R}(\deltasps\mp\Delta B)\bigr],\label{eq:SELRm}
\end{align}
\end{subequations}
corresponding to absorption and emission of magnons by the dot, respectively.
This is the general form for arbitrary $\Delta B$ and $\mu_r$ (the latter enters the Bose function). For investigation of the two driving mechanisms we set either $\Delta B = 0$ ($\Delta\mu$-configuration) or $\mu_L=\mu_R=0$ ($\Delta B$-configuration). 
The product of the spectral- and Bose function, entering~\eqref{eq:SELRp}, is depicted in Fig.~\ref{fig:dB_rates} (left).
While the spectral function introduces the band edges, the Bose function leads to the exponential decay of the rates.
\section{Results\label{sec:results}}
\begin{figure}
\includegraphics[width=0.7\linewidth]{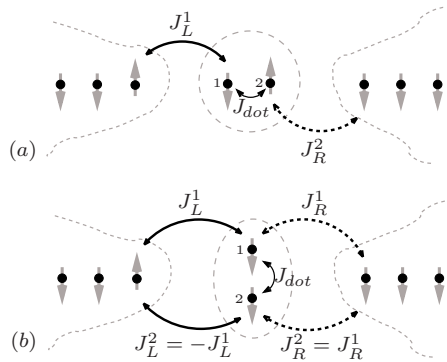}
\caption{\label{fig:coupleconfigs}
The two coupling configurations studied: $(a)$ Serial coupling. $(b)$ Parallel coupling, note the minus sign.}
\end{figure}

First we state some basic principles determining the single magnon transport through the dot, for the case of parallel magnetized reservoirs (the modifications for the case of antiparallel magnetized reservoirs will be stated in section~\ref{sec:antiparallel}).\\  
Tunneling of a magnon out of/into a reservoir involves a spin-flip $\Delta m = \pm 1$ in the dot. Moreover it requires an transition energy above the lower band edges $\omega_{L/R}$ of the corresponding reservoirs.\footnote
{
For the parameters in consideration the upper band edge is of minor importance because there the magnon excitation is already suppressed exponentially.
}
The magnon dispersion relation enters the rates via the spectral function $\Gamma(\omega)$.
At last the tunnel couplings $J^i_r$, which enter the rates as prefactors to the spin matrix elements, impose further restrictions onto the tunneling transitions. Fig.~\ref{fig:coupleconfigs} shows the two coupling configurations studied, which we refer to as \emph{serial} and \emph{parallel coupling}.
For instance these may be realized by two spin\nobreakdash-$1/2$ quantum dots coupled to the reservoirs in series or in parallel, respectively. Particularly for the case of parallel coupling we have examined the influence of a relative phase between the couplings. For vanishing phase no qualitatively new features occur in comparison to the serial coupling case. But for a phase of $\pi$ new features, such as the 2-magnon tunneling, arise. Therefore we restrict the discussion of the parallel coupling to this case, which is depicted in Fig.~\ref{fig:coupleconfigs}(b).\\
Depending on the couplings some states may decouple from either of the reservoirs. E.g., in the case of \emph{parallel coupling} transitions between $\ket{\So}$ and $\ket{\Td}$ can occur over the left reservoir only, while transition between $\ket{\To}$ and $\ket{\Td}$ occur exclusively over the right one; see Fig.~\ref{fig:coupleconfigs}(b).\\
\begin{figure}
\includegraphics[width=\linewidth]{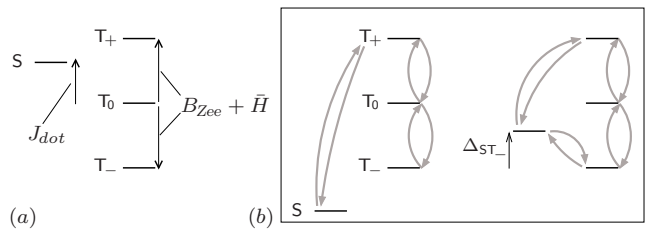}
\caption{\label{fig:eglevelscheme}
$(a)$ The dot level structure for $\ket{\So}$/$\ket{\To}$-eigenstates of $\Hdot$ and  its dependence on the dot magnetic field $\Bzee$ and the intra-dot exchange coupling $\Jdot$. The average $\bar{H}$ of the effective field $H_i$ adds to the homogeneous magnetic field $\Bzee$.
$(b)$ The possible transitions in case of serial coupling and $\ket{\So}$/$\ket{\To}$-eigenstates. Each transition indicated by an arrow can occur over both of the reservoirs. With increasing $\deltaSTd$ beyond the band edge $\omega_{L/R}\geq 0$ the $\So\Td$-transition over the reservoir~$L/R$ becomes possible.}
\end{figure}\noindent

\begin{figure}
\includegraphics[width=0.6\linewidth, height=0.25\linewidth]{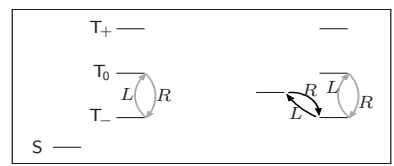}
\caption{\label{fig:egcurrentscheme}
The largest contributions to the spin current~$I_L$, exemplarily for serial coupling and two different level configurations. Darker arrows correspond to larger currents.}
\end{figure}

Fig.~\ref{fig:eglevelscheme}(a) shows the dot level structure, exemplarily for $\ket{\So}$/$\ket{\To}$-eigenstates, i.e.~$\dH = 0$, and its dependence on the dot magnetic field $\Bzee$ and the intra-dot exchange coupling $\Jdot$. The average $\bar{H}= (H_1 + H_2)/2$ of the effective field $H_i$ adds to the homogeneous magnetic field $\Bzee$.
The possible transitions in the case of serial coupling are illustrated in Fig.~\ref{fig:eglevelscheme}(b) [note that transitions can only occur between states where the state with higher energy has also a higher spin quantum number in $z-$direction, since absorption (emission) of a magnon by the dot increases (decreases) energy and spin simultaneously].
In this case, the couplings impose no further restrictions, i.e., if the transition energy between the dot levels is sufficient, the respective transition can occur over both of the reservoirs. \\
In view of the spin current the various transitions contribute differently; see Fig.~\ref{fig:egcurrentscheme}, where only the essential current contributions are shown. Due to the exponential decay of the magnon occupation with energy, large transition energies lead to small current contributions due to suppressed magnon absorption processes (note that current can only occur if there is a closed cycle of magnon absorption and emission processes). Therefore the $\Tu\So$-contribution is left out.
Furthermore the upper levels are generally less occupied and thus play also a minor role in current transport, as long as other `channels' are available, as is the case in Fig.~\ref{fig:egcurrentscheme}.
Since the stationary dot spin is conserved it is sufficient to consider the current $I_L$ over the left reservoir. W.l.o.g.~we choose $\mu_{L/R}$, resp.~$\Delta B$ such, that the spin current flows from the left to the right reservoir. 
Due to its insignificance the backflow from the right to the left is also not depicted in Fig.~\ref{fig:egcurrentscheme}.
%
%
\subsection{Spin current for $\Delta\mu$-configuration\label{sec:deltamu}}
\begin{figure}
\includegraphics[width=\linewidth]{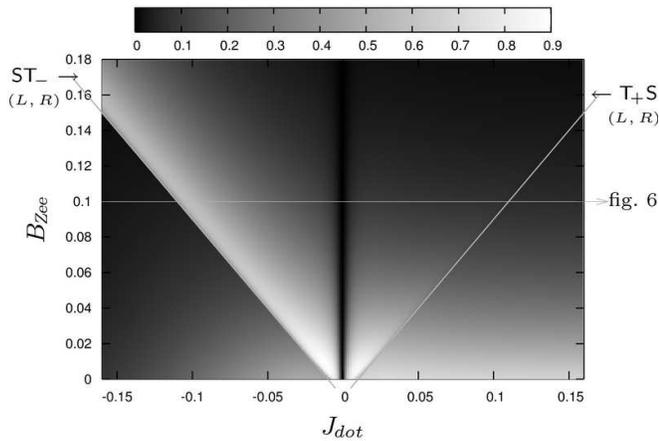}
\caption{\label{fig:Ir_BdJd1}
The spin current $I_L\,(\times 10^4/JS)$ for the $\Delta\mu$-configuration in serial coupling as a function of the dot magnetic field $\Bzee$ and the intra-dot exchange coupling $\Jdot$, both normalized to $JS$. The lines indicate the onsets of transitions.\newline
\qquad\sizeiii[$T=0.05,\;(J^1_L,J^2_L,J^1_R,J^2_R) = ( 3, 0, 0, 1 )\cdot 10^{-2},\;\dH =0.005,\;\bar{H} = 0.01,\;b_1=0.1,\;\mu_{L,R} = (-0.01, -0.3 )$]}
\end{figure}

\begin{figure}
\includegraphics[width=\linewidth]{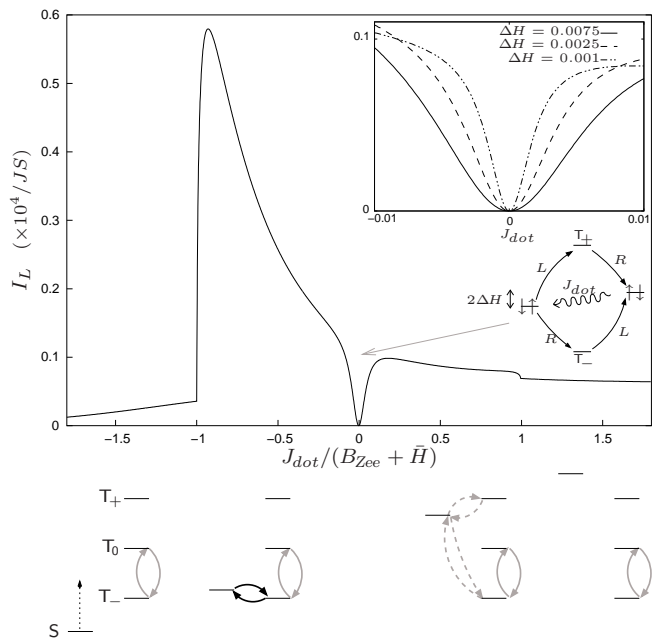}
\caption{\label{fig:Ir_Jd1}The spin current $I_L$ as a function of $\Jdot$ (cross section in \ref{fig:Ir_BdJd1}). The level schemes at the bottom illustrate the main current contributions, compare Fig.~\ref{fig:egcurrentscheme}. Inset: dip for different inhomogeneities $\dH$ and level scheme in the neighborhood of $\Jdot = 0$.\quad\sizeiii[$T=0.05,\;\Bzee = 0.1,\;(J^1_L,J^2_L,J^1_R,J^2_R)= ( 3, 0, 0, 1 )\cdot 10^{-2},\;\dH =0.005,\;\bar{H} = 0.01,\;b_1=0.1,\;\mu_{L,R} = (-0.01, -0.3 )$]
}\end{figure}

{\sffamily Serial coupling.}\quad
For the $\Delta\mu$-configuration the band edges are given by $\omega_L = \omega_R = 0$. Therefore transitions are possible for arbitrary small transition energies.
Fig.~\ref{fig:Ir_BdJd1} shows the spin current for the $\Delta\mu$-configuration in serial coupling as a function of the dot magnetic field $\Bzee$ and the intra-dot exchange coupling $\Jdot$.
With each new channel opened 
the spin current increases. For $\Jdot=0$ the current vanishes, since in this case the whole system separates into two uncoupled parts (left and right); see Fig.~\ref{fig:coupleconfigs}(a). 
The cross section indicated by the arrow is shown in Fig.~\ref{fig:Ir_Jd1}, where the level schemes illustrate the main current contributions. Each time the singlet level crosses one of the triplet levels, a qualitative change of the spin current characteristics occurs.
The outstanding current peak is due to the onset of the $\So\Td$-transition, opening a new current channel which involves the strongly occupied $\Td$-level.\\
The width of the dip at $\Jdot = 0$ is determined by the inhomogeneity~$\dH$ of the effective field $H_i$; see upper inset in Fig.~\ref{fig:Ir_Jd1}. To understand this, consider the $\Jdot = 0$ eigenstates $\ket{\udo}$/$\ket{\duo}$ of $\Hdot$, to which a finite $\Jdot \ll \dH$ may be regarded as a perturbation. The corresponding level scheme is depicted in the lower inset. Here, each transition may exclusively occur over one of the reservoirs only. As a consequence the stationary current vanishes for $\Jdot = 0$, since there exists no path leading back to an arbitrary initial state while at the same time carrying spin from the left to the right reservoir. E.g., on the path $\ket{\duo}\xrightarrow[]{\scriptscriptstyle L}\ket{\Tu} \xrightarrow[]{\scriptscriptstyle R}\ket{\udo}$ a magnon tunnels from left to right. But in order to get back to the initial state, which is necessary to get a stationary current, a magnon must tunnel back again. So effectively the stationary spin current vanishes.
Switching on $\Jdot$ evokes a `coupling' between the $\ket{\udo}$/$\ket{\duo}$-states. Since this coupling opens a new channel between $\ket{\udo}$ and $\ket{\duo}$ without magnon flow, the paths indicated in the inset level scheme give rise to a nonvanishing spin current. The $\Jdot$-mediated coupling becomes important for $\Jdot$ of the order of $\dH$, which measures the splitting between the two levels. Thus $\dH$ determines the width of the current dip.\\
For $\abs{\dH} \ll \abs{\min\{J^i_r\}}$, particularly $\dH =0$, a narrow dip still remains. In this case the eigenstates are $\So$/$\To$-like also in the neighborhood of $\Jdot = 0$. Then, the suppression of the spin current in the dip is due to a destructive interference between equivalent current carrying paths, e.g.~the $\So\Td$- and $\To\Td$-paths. Thus, here the nondiagonal matrixelement $\pst_{\So\To}$ has to be taken into account, i.e.~only the full kinetic equation yields correct results in the neighborhood of $\Jdot =0$. \\
Tuning further up to $\Jdot > \Bzee$, the spin current finally saturates in the contribution of the $\To\Td$-transition, while the $\So$-level occupation gets negligible.

\begin{figure}
\includegraphics[width=\linewidth]{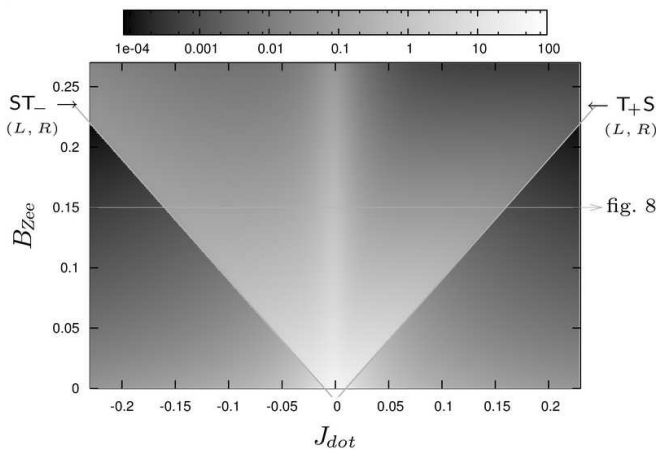}
\caption{\label{fig:Ir_BdJd2}
The spin current $I_L\,(\times 10^5/JS)$ for the $\Delta\mu$-configuration in parallel coupling as a function of $\Bzee$ and $\Jdot$. In contrast to serial coupling here a peak occurs for $\Jdot = 0$ (note the logarithmic scale).\newline
\qquad\sizeiii[$T=0.05,\;(J^1_L,J^2_L,J^1_R,J^2_R) = ( 1, -1, 2, 2 )\cdot 10^{-2},\;\dH =0.005,\;\bar{H} = 0.01,\;b_1=0.1,\;\mu_{L,R} = (-0.01, -0.3 )$]}
\end{figure}

\begin{figure}
\includegraphics[width=\linewidth]{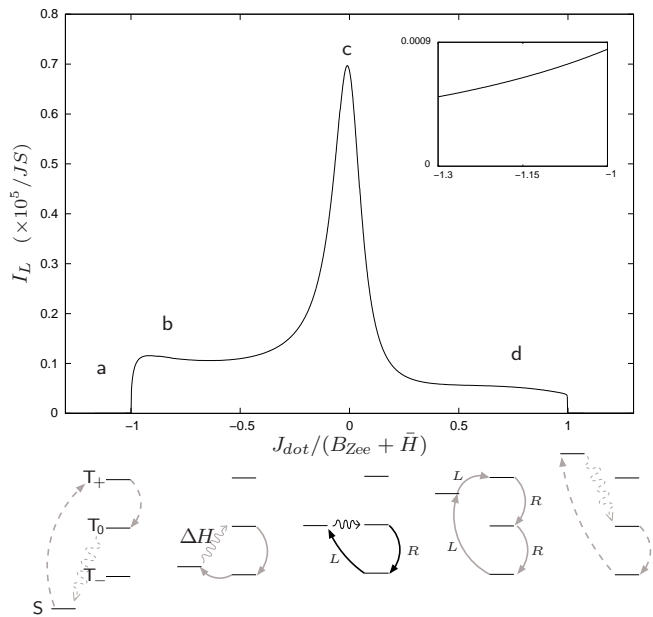}
\caption{\label{fig:Ir_Jd2} 
The spin current $I_L$ as a function of $\Jdot$ (cross section in \ref{fig:Ir_BdJd2}). Note that here $\dH$ plays the role of the perturbation. The inset shows the magnified sector~$(a)$.
\quad\sizeiii[$T=0.05,\;\Bzee = 0.15,\;(J^1_L,J^2_L,J^1_R,J^2_R) = ( 1, -1, 2, 2 )\cdot 10^{-2},\;\dH =0.005,\;\bar{H} = 0.01,\;b_1=0.1,\;\mu_{L,R} = (-0.01, -0.3 )$]
}\end{figure}

{\sffamily Parallel coupling.}\quad
Fig.~\ref{fig:Ir_BdJd2} shows the spin current for parallel coupling as a function of $\Bzee$ and $\Jdot$.
For $\Jdot = 0$ instead of a dip, now a \emph{peak} occurs in the spin current. In contrast to the current peak in Fig.~\ref{fig:Ir_Jd1} this peak cannot be ascribed to the onset of a transition. To understand this we discuss the current along the cross-section indicated; see Fig.~\ref{fig:Ir_Jd2}. 
Again, the level schemes illustrate the largest contributions to the spin current. 
The difference between serial and parallel coupling is, that in the case of the latter the transitions are more restricted [as mentioned in the context of Fig.~\ref{fig:coupleconfigs}(b)].
Transitions involving $\ket{\So}$ ($\ket{\To}$) may occur exclusively over the left (right) reservoir here. Therefore a stationary spin current requires a sequential tunneling of \emph{two} magnons, e.g.~on the path 
$\ket{\Td}\xrightarrow[]{\scriptscriptstyle L}\ket{\So}\xrightarrow[]{\scriptscriptstyle L}\ket{\Tu}\xrightarrow[]{\scriptscriptstyle R}\ket{\To}\xrightarrow[]{\scriptscriptstyle R}\ket{\Td}$, as illustrated in the level scheme for sector~$(d)$.\\ 
The current peak~$(c)$ appears for $\Jdot \lesssim \dH$ since then the inhomogeneity $\dH$ becomes essential by `coupling' the $\ket{\So}$/$\ket{\To}$-states and thus introducing a further channel which enables 1-magnon tunneling; see the central level scheme. 
In this way the otherwise inevitable 2-magnon tunneling is cut short, resulting in an increased current [compared to the 1-magnon tunneling peak~$(c)$ the current in sector~$(d)$ is exponentially decreased by a factor of $\exp(-\beta\Delta_{\To\Td})$].\\
Unlike in sector $(d)$ the $\dH$-mediated coupling still takes influence in sector $(b)$, as illustrated in the according level scheme.
While in the former case the $\So\Td$-transition is the `bottleneck', which cannot be short-circuited, in the latter case the restricting $\Tu\So$-transition is cut short indeed.\\
%
Besides these qualitative arguments it shall be mentioned that the quantitative computation of the spin current is performed by setting up the kinetic equation in terms of the eigenbasis of $\Hdot$. The influence of nondiagonal matrixelements between the two states with $m=0$, which gets most important for a small level splitting, has to be considered too. For $\Jdot = 0$ the eigenstates are  $\ket{\udo}$/$\ket{\duo}$ while their minimal splitting is given by $\dH$.
Since the inhomogeneity $\dH$ cannot vanish for the $J^i_r$ in parallel coupling, the states are never degenerate and thus a destructive interference is largely suppressed. However a small current correction still remains, resulting in a reduced peak height. Thus, in order to account for this correction, the full kinetic equation has to be considered instead of the classical master equation. 

\subsection{Spin current for $\Delta B$-configuration\label{sec:deltaB}}
\begin{figure}
\includegraphics[width=\linewidth]{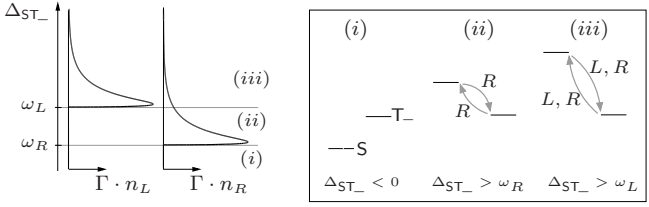}
\caption{\label{fig:dB_rates}
(Left) The product of spectral- and Bose function, which determines the energy dependence of the rates [compare~\eqref{eq:SELR}], for the left and right reservoir as a function of transition energy $\deltaSTd$. The band edges are given by $\omega_{L/R} = B_0 \pm \Delta B$.
(Right) The onset of transitions illustrated by means of the $\So\Td$-transition; for clarity the upper levels are left out. When $\deltaSTd$ exceeds $\omega_{R/L}$ the transition over the right/left reservoir becomes possible.}
\end{figure}

In the case of the $\Delta B$-configuration a magnetic field gradient drives the spin current. Here the minimal excitation energies in the left and right reservoir differ, since the band edges are shifted by the magnetic field gradient: $\omega_{L/R} = B_0 \pm \Delta B$. Therefore, in contrast to the $\Delta\mu$-configuration, the transitions over the left and right reservoir set in for \emph{different} dot transition energies which leads to additional features in the spin current. 
The energy dependence of the rates entering the kinetic equation, particularly the shift of the band edges, is shown in Fig.~\ref{fig:dB_rates}. It also illustrates the successive onset of the transitions over the right and left reservoir, exemplarily for the $\So\Td$-transition.

\begin{figure}
\includegraphics[width=\linewidth]{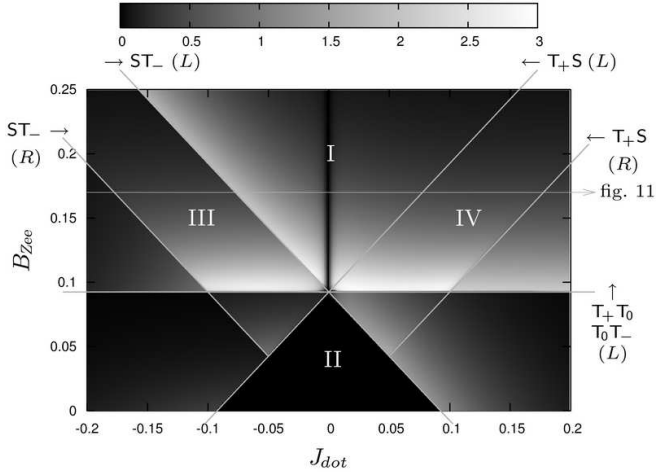}
\caption{\label{fig:Ir_BdJd3}
The spin current $I_L\,(\times 10^5/JS)$ for the $\Delta B$-configuration in serial coupling as a function of $\Bzee$ and $\Jdot$. \newline
\quad\sizeiii[$T=0.05,\;(J^1_L,J^2_L,J^1_R,J^2_R) = ( 2, 0, 0, 1 )\cdot 10^{-2},\;\dH =2.5\cdot 10^{-3},\;\bar{H} = 7.5\cdot 10^{-3},\;b_1=0.1,\;\Delta B = B_0 =0.05$]}
\end{figure}

\begin{figure}
\includegraphics[width=\linewidth]{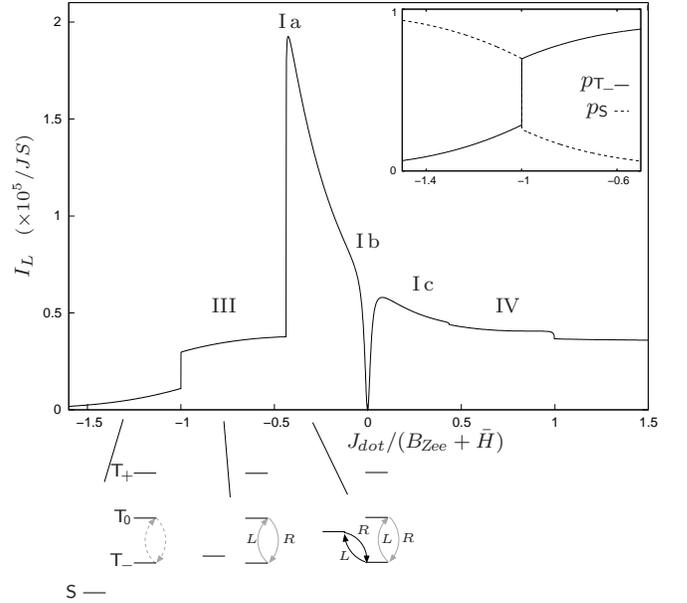}
\caption{\label{fig:Ir_Jd3}The spin current $I_L$ as a function of $\Jdot$ (cross section in Fig.~\ref{fig:Ir_BdJd3}). In sector~$(\textup{III})$ the $\So\Td$-transition over the right reservoir sets in. Although no new current channel is opened, the current increases due to a swap of occupation between the $\So$- and $\Td$-level; compare the inset showing the occupation of the $\So$- and $\Td$-level in this sector.
\quad\sizeiii[$T=0.05,\;\Bzee = 0.17,\;(J^1_L,J^2_L,J^1_R,J^2_R) = ( 2, 0, 0, 1 )\cdot 10^{-2},\;\dH =2.5\cdot 10^{-3},\;\bar{H} = 7.5\cdot 10^{-3},\;b_1=0.1,\; \Delta B= B_0 =0.05$]
}\end{figure}

{\sffamily Serial coupling.}\quad Fig.~\ref{fig:Ir_BdJd3} shows the spin current for the $\Delta B$-configuration in serial coupling.
Since we choose $\Delta B = B_0$ here, it follows that $\omega_R = 0$. However, a different choice does not lead to relevant modifications. 
Again the onsets of transitions are indicated by lines. 
For $\Jdot \lesssim \dH$ deviations from these lines occur due to the influence of the inhomogeneity $\dH$, which shifts the eigenenergies of $\Hdot$ non-linearly.\\ 
In sector~$(\textup{I})$ the current is basically the same as for the case of the $\Delta\mu$-configuration, compare Fig.~\ref{fig:Ir_BdJd1}.  
Beyond this similarity new features arise, specific to the $\Delta B$-configuration.
In the lower half of Fig.~\ref{fig:Ir_BdJd3} an area of zero current occurs, denoted as sector $(\textup{II})$. The left reservoir is completely decoupled here since all transitions involving it are disabled (all the transition energies lie below the left band edge). Furthermore there are additional steps in the sectors $(\textup{III})$ and $(\textup{IV})$. To understand these, consider Fig.~\ref{fig:Ir_Jd3} which shows the cross section indicated.
%
%
In sector~$(\textup{III})$ the onset of the $\So\Td$-transition over the right reservoir [compare Fig.~\ref{fig:dB_rates}(b)] does not open a new current channel but leads to a swap of occupation between the $\So$- and $\Td$-level; see inset. Thus the increase of the spin current is due to the enhanced $\Td$-occupation. This is revealed by the fact that the $\Td$-occupation is effectively mapped onto the current (compare inset). An analog argumentation holds for the step in sector~$(\textup{IV})$.\\
In contrast to this, the peak~$(\textup{I\,a})$ arises since a new (resonant) current channel is opened by the onset of the $\So\Td$-transition over the left reservoir. This is completely analog to the corresponding peak in the case of the $\Delta\mu$-configuration (compare Fig.~\ref{fig:Ir_Jd1}). For sectors $(\textup{I\,b})$ and $(\textup{I\,c})$ there is no qualitative difference as well since all transitions are possible over both reservoirs, as is the case for the $\Delta\mu$-configuration too.\\ 
{\sffamily Parallel coupling.}\quad Analog to the case of the $\Delta\mu$-configuration (compare Fig.~\ref{fig:Ir_BdJd1} and \ref{fig:Ir_BdJd2}) for $\Jdot =0$ the dip is replaced by a peak in sector $(\textup{I})$. Nevertheless the sector $(\textup{II})$ of zero current, as also the steps in sectors $(\textup{III})$ and $(\textup{IV})$ remain as for the case of serial coupling.

\begin{figure}
\includegraphics[width=\linewidth]{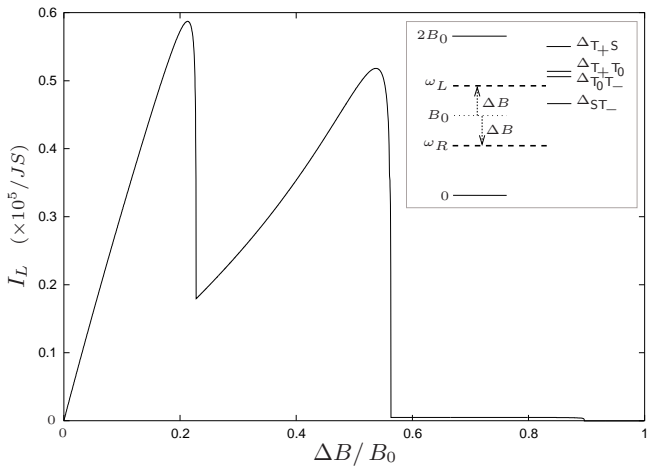}
\caption{\label{fig:Ir_dB1} The spin current $I_L$ as a function of the magnetic field gradient $\Delta B$. Inset: the transition energies and the band edges $\omega_{L/R} = B_0 \pm \Delta B$. When a band edge (here $\omega_L$) exceeds a transition energy the corresponding transition sets out. The splitting between $\Delta_{\Tu\To}$ and $\Delta_{\To\Td}$ is approximately given by $2\dH^2/\Jdot = 3.1\cdot 10^{-4}$, which cannot be resolved on the scale shown.
\quad\sizeiii[$T=0.05,\;\Jdot= -0.04,\;\Bzee = 0.18,\;(J^1_L,J^2_L,J^1_R,J^2_R) = ( 2, 0, 0, 1 )\cdot 10^{-2},\;\dH =2.5\cdot 10^{-3},\;\bar{H} = 7.5\cdot 10^{-3},\;b_1=0.1,\;B_0 =0.12$]
}\end{figure}

\begin{figure}
\includegraphics[width=\linewidth]{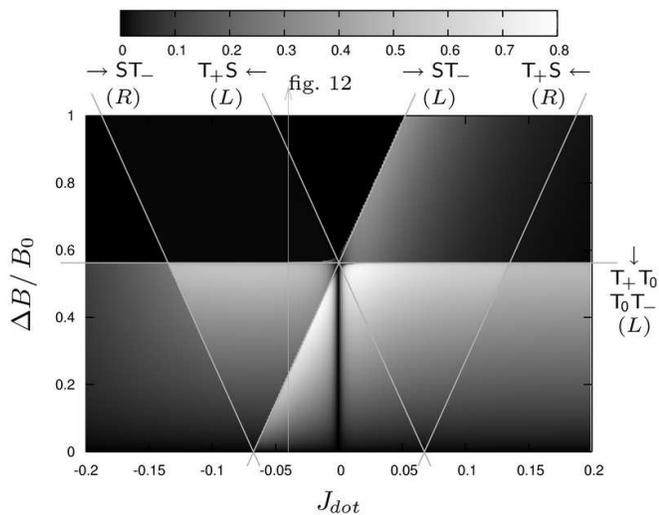}
\caption{\label{fig:Ir_dBJd4}
The spin current $I_L\,(\times 10^5/JS)$ for the $\Delta B$-configuration in serial coupling as a function of the \emph{magnetic field gradient} $\Delta B$ and $\Jdot$. 
\quad\sizeiii[$T=0.05,\;\Bzee = 0.18,\;(J^1_L,J^2_L,J^1_R,J^2_R) = ( 2, 0, 0, 1 )\cdot 10^{-2},\;\dH =2.5\cdot 10^{-3},\;\bar{H} = 7.5\cdot 10^{-3},\;b_1=0.1,\;B_0 =0.12$]
}
\end{figure}

Finally, for the case of serial coupling, we discuss the spin current as a function of the \emph{magnetic field gradient} $\Delta B$; see Fig.~\ref{fig:Ir_dB1}, where the dot parameters $\Bzee$ and $\Jdot$ are fixed. 
For small $\Delta B$ the current increases linearly. Then, due to the shift of the band edges, the transitions over the left reservoir set out successively (see inset), each resulting in a drop of the current. When the left reservoir finally decouples, the current vanishes completely. \\
The shape of the current as function of $\Delta B$ depends strongly on the dot level structure determining which transitions are involved in which order. Fig.~\ref{fig:Ir_dBJd4} shows exemplarily the current as a function of $\Delta B$ and $\Jdot$. 
Once again with each onset of a transition the current increases. 

\subsection{Antiparallel magnetized reservoirs\label{sec:antiparallel}}
\begin{figure}
\includegraphics[width=\linewidth]{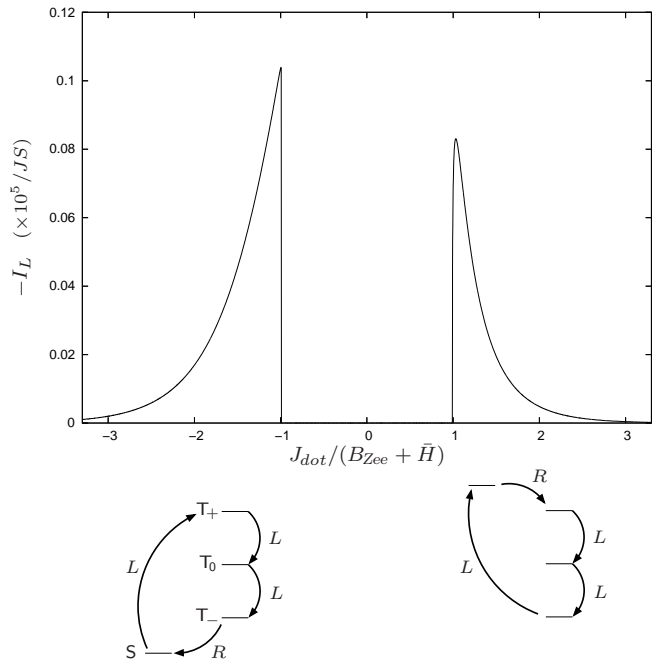}
\caption{\label{fig:Ir_anti}The spin current $I_L$ as a function of $\Jdot$ for the case of antiparallel magnetized reservoirs in serial coupling. When the $\So$-level lies between the $\Td$- and $\Tu$-levels the current is suppressed due to decoupling of the right reservoir. In the remaining sectors the current is carried by the transport cycles depicted in the level schemes. \quad\sizeiii[$T=0.05,\;\Bzee = 0.1,\;(J^1_L,J^2_L,J^1_R,J^2_R)= ( 3, 0, 0, 1 )\cdot 10^{-2},\;\dH =0.01,\;\bar{H} = 0.005,\;b_1=0.1,\;\mu_{L,R} = (-0.03, -0.03 )$]}
\end{figure}
So far we have considered the reservoirs to be magnetized in the same direction. In the following we briefly discuss the case of two antiparallel magnetized reservoirs, where w.l.o.g.~the ground state of the right reservoir is flipped upwards. The real-time formalism used is also capable to deal with such a system. This is simply achieved by exchanging the magnon creation and annihilation operators for the right reservoir in~\eqref{eq:Ham_r}, \eqref{eq:HamT} and~\eqref{eq:hatI}.\footnote
{This results in exchanging $n^+_r \leftrightarrow n^-_r$ and reversing the sign of $\omega\rightarrow -\omega$ in the resolvent of the expression~\eqref{eq:sigma} [the latter involves a sign reversal of the arguments of the spectral- and Bose functions in~\eqref{eq:SELR}]. 
}\\
In this antiparallel configuration a spin current flows even without an additional driving mechanism, such as a magnetic field gradient applied, since the reservoir magnetizations tend to equalize by exciting magnons in both of the reservoirs.
Fig.~\ref{fig:Ir_anti} shows exemplarily the spin current as a function of~$\Jdot$ for the case of serial coupling. In view of the basic transport principles the main change is, that transitions over the right reservoir can now only occur between states where the state with higher energy has the lower spin quantum number in $z-$direction, since the magnons in the right reservoir carry a `spin-down'.\\
The spin current $I_L$ is negative since `spin-ups' are entering the left reservoir, in order to equalize the magnetizations.
The current is suppressed, when the singlet level lies between the two outer triplet levels, since then the right reservoir is completely decoupled. In the remaining sectors a non-vanishing spin current is carried by magnon tunneling processes which involve all dot levels, as indicated in the level schemes in Fig.~\ref{fig:Ir_anti}. Since these processes require a magnon to be initially excited in the left reservoir (long upward arrow in the level schemes), the current decays exponentially with increasing $\abs{\Jdot}$. Furthermore, due to this the current vanishes with vanishing temperature.\\ 
We conclude that also in this antiparallel configuration the spin current can be switched by tuning the intra-dot parameters $\Bzee$ and $\Jdot$, e.g.~by simply evoking a $\So\Td$-crossing.
\subsection*{Conclusions}
In summary, our investigation of the proposed spin quantum dot system revealed a rich structure of the spin current as a function of the intra-dot parameters $\Bzee$ and $\Jdot$. Due to the involved large current variations, a switching of the current is possible in principle. \\
For the case of \emph{parallel magnetized reservoirs} the influence of the dot-reservoir couplings on the spin transport becomes apparent by opposing the serial and parallel coupling configurations. The restriction of the transitions in the case of parallel coupling leads to drastically different transport characteristics, which originate from the sequential 2-magnon tunneling.
However, due to the differing signs of the dot-reservoir couplings, the parallel coupling configuration represents rather an extreme case compared to the serial coupling configuration which suggests itself. 
It shall be mentioned that small deviations from the `pure' coupling configurations discussed here, do not lead to qualitative modifications of the spin current results.\\
The concept of a magnetochemical potential as sum of the chemical potential and the magnetic field, analog to the electrochemical potential in the electronic case, is deficient.
The magnon chemical potential controls the magnon occupation, whereas an applied magnetic field gradient shifts the magnon band edges while leaving the overall magnon occupation unchanged. As we have shown, both mechanisms, the differing left and right magnon chemical potentials as also the magnetic field gradient, yield different results for the spin current. This is due to the fact that the magnon dispersion relation, particularly the band edges, play an essential role in spin transport, in contrast to the case of electronic transport, which primarily takes place between the electrochemical potentials.

In the case of \emph{antiparallel magnetized reservoirs} the main features of the spin current can be explained by considering slightly modified spin transport principles. As we have shown, here spin current switching is possible too.

We remark, that in those sectors where the spin current carried by single magnon tunneling is exponentially suppressed, a cotunneling of magnons, analog to its electronic pendant, may yield non-negligible contributions to the spin current. However, the current suppression due to magnon dispersion effects should remain untouched.\\
%
%
%
%

We gratefully acknowledge discussions with F.~Meier, J.~Martinek, D.~Loss and H.~Capellmann.

\appendix*
\section{}
\begin{figure*}\centering
\includegraphics[width=0.9\linewidth]{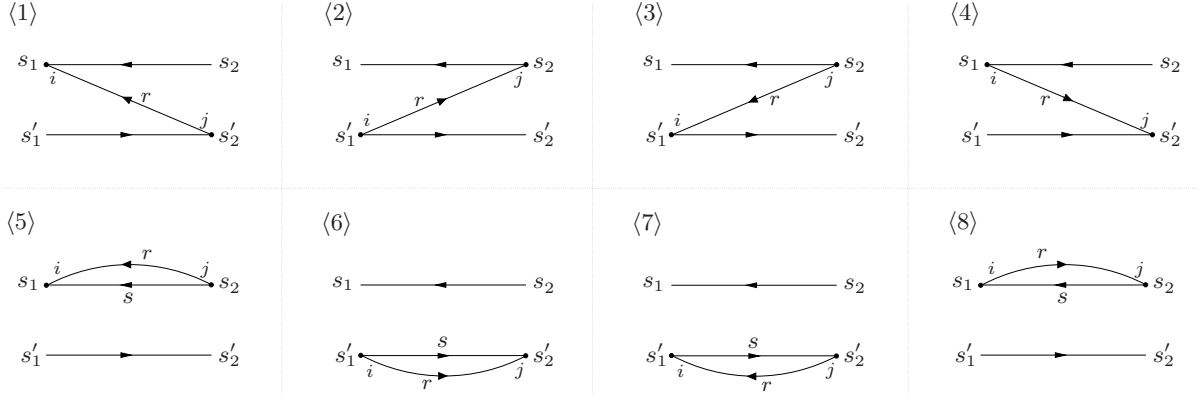}
\caption{\label{fig:diagrams} 
The diagrams contributing to $\Sigma(\iop)$ and $\Sigma_{I_r}(\iop)$ to lowest order in tunneling.}
\end{figure*}

The diagrams contributing to the kernels $\Sigma(\iop)$ and $\Sigma_{I_r}(\iop)$ to lowest order in tunneling are shown in Fig.~\ref{fig:diagrams}. 
Thereby, in matrix representation the kernels are given by
\begin{align}\Sigma(\iop)_{\si\sip,\sii\siip} &= \sum_{r=L,R}\,\sum_{d=1}^{8}\:\langle d\rangle\\
\Sigma_{I_r}(\iop)_{\si\sip,\sii\siip} &= \langle 1 \rangle + \langle 2 \rangle + \langle 5 \rangle +\langle 6 \rangle\,.
\end{align}
By defining the auxiliary function
\begin{equation}\label{eq:sigma}
\sijrp{i}{j}{r}{p}(\Delta E) = \frac{\imag}{2\pi}\int\nts d\omega\:\frac{\Gamma(\omega)\,n_r^p(\omega)}{\Delta E -\bar{B}_r -\omega + \iop}
\end{equation}
where $p \in \{+,-\}$, $n_r^+(\omega) = n_r(\omega)$, $n_r^-(\omega) = 1 + n_r(\omega)$ and $\bar{B}_{L/R} = \pm \Delta B$, the expressions corresponding to the diagrams with left running reservoir contractions are given as follows:
%
\begin{align*}
\langle 1 \rangle &= \sum_{i,j}\, \bran{\si} J^i_r s_{i}^+ \ketn{\sii}\bran{\siip} J^j_r s_{j}^- \ketn{\sip}
\;\sijrp{i}{j}{r}{+}(\Delta_{\sip\sii})\,,\\ 
\langle 3 \rangle &= \sum_{i,j}\, \bran{\si} J^j_r s_{j}^-\ketn{\sii}\bran{\siip} J^i_r s_{i}^+ \ketn{\sip}
\;\sijrp{i}{j}{r}{-}(\Delta_{\siip\si})\,,\\ 
\langle 5 \rangle &= -\sum_{i,j}\sum_{s}\delta_{\sip\siip}\, \bran{\si} J^i_r s_{i}^+ \ketn{\s}\bran{\s} J^j_r s_{j}^- \ketn{\sii}\,
\sijrp{i}{j}{r}{-}(\Delta_{\sip\s})\,,\\
\langle 7 \rangle &= -\sum_{i,j}\sum_{s}\delta_{\si\sii}\, \bran{\siip} J^j_r s_{j}^- \ketn{\s}\bran{\s} J^i_r s_{i}^+ \ketn{\sip}\,
\sijrp{i}{j}{r}{+}(\Delta_{\s\si})\,. 
\end{align*}
%
%
Each of the remaining diagrams $\langle d \rangle$ with right running contraction is obtained by taking the complex conjugate and interchanging the primed by unprimed indices (and vice versa) in the expression for the left running diagram $\langle d-1 \rangle$. E.g.:
\begin{equation*}
\langle 2 \rangle = \biggl[\sum_{i,j}\, \bran{\sip} J^i_r s_{i}^+ \ketn{\siip}\bran{\sii} J^j_r s_{j}^- \ketn{\si}
\;\sijrp{i}{j}{r}{+}(\Delta_{\si\siip})\biggr]^*.
\end{equation*}


\end{document}